# Wavelet Image Restoration Using Multifractal Priors

Karl Young, John Kornak, Eric Friedman

Department of Epidemiology and Biostatistics, University of California San Francisco, Mission Hall: Global Health & Clinical Sciences Building
550 16th St, 2nd floor, Box #0560

Running Title:    Wavelet Image Enhancement Using Multifractal Priors

Address Correspondence to:
John Kornak, PhD
Professor in Residence
Head of Data Science Program
Department of Epidemiology and Biostatistics
University of California, San Francisco
Mission Hall: Global Health & Clinical Sciences Building
550 16th St, 2nd floor, Box #0560
San Francisco, CA 94158-2549
Tel: 415-514-8028
Fax: 415-514-8150
Email: john.kornak@ucsf.edu

# Abstract

Bayesian image restoration has had a long history of successful application but one of the limitations that has prevented more widespread use is that the methods are generally computationally intensive. The authors recently addressed this issue by developing a method that performs the image enhancement in an orthogonal space (Fourier space in that case) which effectively transforms the problem from a large multivariate optimization problem to a set of smaller independent univariate optimization problems. The current paper extends these methods to analysis in another orthogonal basis, wavelets. While still providing the computational efficiency obtained with the original method in Fourier space, this extension allows more flexibility in adapting to local properties of the images, as well as capitalizing on the long history of developments for wavelet shrinkage methods. In addition, wavelet methods, including empirical Bayes specific methods, have recently been developed to effectively capture multifractal properties of images. An extension of these methods is utilized to enhance the recovery of textural characteristics of the underlying image. These enhancements should be beneficial in characterizing textural differences such as those occurring in medical images of diseased and healthy tissues. The Bayesian framework defined in the space of wavelets provides a flexible model that is easily extended to a variety of imaging contexts.

# Introduction

Pixel-based Bayesian image restoration has a long and successful history [1] described in [2] and the references cited therein. Further improvements to the methods constitute on ongoing area of active research [3]. In particular the authors recently developed a method to address one of the long-standing issues in Bayesian image processing, i.e. the fact that the methods can be computationally intensive [4]. This computational intensity can be attributed to framing the image restoration problem in terms of the posterior distribution over the set of pixels in the image. Due to pixel correlations in the images, obtaining summary images from the posterior distribution (such as the Maximum a-posteriori, MAP, estimated image) amounts to performing a large multivariate optimization or sampling procedure.

The proposal of the authors was to move the determination of the posterior distribution to an orthogonal domain; that was Fourier space in the initial proposal. It was argued that in Fourier space a reasonable approximation would be to treat the Fourier modes as independent, thus rendering the problem of obtaining an optimized image estimate from the posterior distribution, as one of obtaining a product of independent optimizations. This in turn greatly reduces the computational burden by transforming the problem from a multivariate optimization to a set of independent single variable optimizations and one that in addition is trivially parallelizable.

The authors demonstrated the effectiveness of the methods, including the appropriateness of the approximation of independence across Fourier space locations, by producing restored images that were comparable to that produced by more standard Bayesian image restoration methods based on the MAP estimator or Markov Chain Monte Carlo (MCMC) based image summaries, including exhibiting similar covariance structure to models developed in the original image space [4].

One of the primary components of the methods proposed, was to generate what the authors referred to as parameter functions. These are functions over the orthogonal space that parametrize the prior distribution at each point in that space and represent prior information about the image. For instance, if the Fourier spectra of a large number of images is examined, it is found that typically most of the power lies in the large wavelength Fourier modes representing the average values over large regions in the image. Values at the shorter wavelength modes are generally small amplitude fluctuations that represent noise. So, a reasonable parameter function for the prior mean in Fourier space for typical images would be a power law, $ak^{-b}$ where $k$ represents the Euclidean distance of the Fourier space coordinates from the origin (representing zero spatial frequency in each of the $x$ and $y$ directions). Assuming that the prior distribution at each Fourier space location is well approximated as a Gaussian then a reasonable parameter function to choose for the variance is one that is proportional to the mean. That choice gives more weight to the likelihood distribution (the data) for long wavelength Fourier modes where the signal to noise ratio is expected to be large and more weight to the prior distribution at short wavelength Fourier modes that are expected to be dominated by noise, thus effectively implementing a smoothing prior. While neither performing image restoration in the wavelet domain, nor doing so in a Bayesian context itself is novel, the degree of flexibility provided by modeling parameter functions in the orthogonal basis, e.g. Fourier or wavelet, provide for a new perspective on the problem in the experience of the authors.

The authors suggested, then demonstrated, that the focus on parameter functions provided the flexibility for generating image-type specific prior models [4]. For example, if one is attempting to restore a particular class of images, e.g. tissue specific medical images generated by a specific imaging protocol, one can generate empirical parameter functions by averaging over a large number of images from the given class. A current project using these methods is focused on the generation of enhanced MRI perfusion images, using parameter functions generated from co-registered and standard-of-care Fludeoxyglucose ($^{18}$F) Positron-Emission Tomography (FDG-PET) imaging. This takes advantage of the fact that the information content across multiple PET images can be used to enhance the higher-noise perfusion images, and thereby effectively enhances the resolution of the restored perfusion images.

The work discussed in this paper, Wavelet Imaging with Multifractal Priors (*WIMP*) began by investigating whether the successful application of the restoration methods in Fourier space could be extended to other orthogonal spaces. In particular, given the extensive literature on image restoration using wavelets, the use of wavelet bases was expected to provide advantages in characterizing location specific properties of the image. This can be difficult using the Fourier basis, given the global nature of the Fourier basis functions in image space. Conversely a concern might be how well the independence of modes assumption would transfer to other orthogonal bases such as wavelets. Strictly speaking, from a functional analysis point of view the independence assumption is appropriate, as the basis functions for orthogonal transformations such Fourier and wavelet transformations are always independent. But for approximation of images at moderate resolutions effectively relying on a projection onto a limited number of basis functions, this is a potential issue.

How to generate appropriate parameter functions in the wavelet domain might seem less intuitive than doing so in the Fourier domain, given the typical expectations about which Fourier modes are dominated by noise. Thus, properties such as scaling behavior would seem less universal for wavelet bases than for the Fourier basis, and consequently harder to model in the wavelet domain. Nonetheless, in early tests, it was found that assuming a decreasing power law for finer wavelet scales did result in effective image restoration.

It turns out that methods developed to implement wavelet shrinkage [5] [6] provide a particularly effective means for generating parameter functions that take advantage of properties of wavelet bases such as sparsity of representation. But note that in the case of the Bayesian imaging methods discussed here, while sparsity of representation can provide certain advantages, it is the assumption of the independence of modes that is still the dominant factor in the development of a computationally efficient restoration method. Also note that while the flexibility of the methods allows for the incorporation of shrinkage methods and provides for convenient comparison of the proposed method with current state of the art wavelet shrinkage-based restoration methods, incorporating shrinkage methods is only one of many ways to generate parameter functions in the wavelet domain.

Since *WIMP* is aimed at application to domains such as medical imaging, advantage is taken of the fact that the use of wavelets has been shown to be effective at characterizing multifractal characteristics of images [7]. And multifractal analysis has been shown to be effective in a number of applications to pattern recognition, texture analysis, and segmentation in medical imaging [8]. Therefore, in addition to the incorporation of wavelet shrinkage for Bayesian restoration, methods for using wavelets to characterize multifractal geometry in images [7] [9] are also applied to the generation of appropriate wavelet domain parameter functions.

# Methods

The general procedure for implementing an image restoration algorithm using the Bayesian imaging in orthogonal space framework is to first choose likelihood and prior distributions, based on the nature of the images and the methods by which they were acquired, then specify appropriate parameter functions for the prior distribution over the indices of the orthogonal space.

The effectiveness of the restoration algorithm can be sensitive to these choices, as there can be competing demands governing the choices. An example is that it is preferable to work with a simple, closed form posterior distribution which can be guaranteed when the prior distribution is conjugate to the likelihood distribution [10]. But sometimes the image noise, characterized by the likelihood distribution, or the choice of an appropriate prior distribution, suggest choices of the likelihood and prior distributions that aren't conjugate. If a closed form for the posterior isn't available but, e.g., a MAP value for each of the points in the orthogonal space is desired, optimization over the product of the likelihood and prior can be performed. But there are cases in which the data value, as likelihood mean, and prior mean value are widely separated such that there is only tiny overlap between the prior and likelihood, to the extent that it can lead to numerical instabilities for the optimization.

An example of when a non-conjugate prior is needed occurs when noise is Gaussian in image space but the Bayesian analysis is performed by working in Fourier space. Specification of the parameter mean is most naturally performed as the function of the modulus of the Fourier space coordinates. But if the distribution in image space is Gaussian then the distribution over real and imaginary components in Fourier space is Gaussian, leading to the distribution over the modulus being Rician [11]. In most cases a Gaussian likelihood is an appropriate choice, but the Rican is not conjugate to the Gaussian. The authors found ways around this issue [4], both by using approximations for the Rician distribution and utilizing the fact that in many situations a Gaussian distribution suffices, representing a simple and often efective example of an approximation of the Rician distribution. We note here that the choice of parameter function is often much more important than the choice of likelihood function.

For the situation in this paper, for which the orthogonal space is taken to be a wavelet space, advantage can be taken of the fact that Gaussian noise in image space transforms to Gaussian noise for the wavelet coefficients [12], as in the Fourier space. We note here that there is a wide range of choices for the wavelet basis and that the combination of image characteristics and distribution functions could partially determine the most appropriate choice. At this point we ignore that detail and discuss the algorithm without regard to the choice of wavelet basis and defer that discussion until later.

Provided that the noise in the original image space is well approximated by Gaussian noise, one can choose Gaussian distributions for both the likelihood and prior distributions resulting in a Gaussian posterior and obtain a simple formula for the MAP estimate based on the conjugate Bayes solution. We assume this to be the case for the examples studied in this paper and use of the form for the MAP estimate turns out to be convenient for emulating both wavelet shrinkage and multifractal estimation in the appropriate parameter domains.

In addition to the choice of likelihood and prior distributions, the parameter functions specifying the variation in the parameters of the prior distribution over the wavelet domain remains. For the case of a Gaussian prior this means the choice of prior mean and prior variance parameter functions needs to be made. In preliminary studies, it was found that a somewhat simplistic choice, similar to the choice for the Fourier basis, of a decreasing power law in wavelet scale, but independent of position, for the prior mean, and a prior variance proportional to the prior mean, was led to improved restored images. But as stated above, for the current paper the goal is to explore the efficacy of attempting to combine Bayesian image analysis in wavelet space with the demonstrated benefits of wavelet shrinkage and multifractal approximation methods.

As a preliminary to specifying parameter functions that emulate wavelet shrinkage and multifractal estimation in the appropriate domains, we base the method on MAP estimation of the wavelet coefficients. The MAP estimate for the mean of the Gaussian posterior at position $u$ and scale $s$ is:

$$M(u,s) = \frac{\frac{W_l(u,s)}{\sigma_l^2} + \frac{W_p(u,s)}{\sigma_p^2(u,s)}}{\frac{1}{\sigma_l^2} + \frac{1}{\sigma_p^2(u,s)}} \quad (1)$$

where $W_l(u,s)$ is the wavelet coefficient at position $u$ and scale $s$, from the wavelet transform of the original image, i.e. the likelihood mean. $\sigma_l^2$ is the mean square of the noise estimated via the method described in [5]: used as the likelihood variance. $W_p(u,s)$ is the value of the prior mean parameter function at position $u$ and scale $s$, and $\sigma_p^2(u,s)$ is the value of the prior variance parameter function at position $u$ and scale $s$.

To specify the prior mean and variance parameter functions, first consider the mean square wavelet coefficients at each scale, averaged over position:

$$W_s^2 = \frac{1}{N_s} \sum_u W_l^2(u,s) \quad (2)$$

where $N_s$ is the number of position coefficients at scale $s$. $W_s^2$ will be used for obtaining the *BayesShrink* [6], and multifractal [9] thresholds.

To utilize wavelet shrinkage properties, note that the basic idea is to capitalize on the sparse representation of typical signals in the wavelet domain. As a result, it is assumed that the large number of coefficient values that are less than a predetermined threshold value represent noise and can be zeroed, and the few coefficients that are greater than the threshold, represent signal, but should "shrink" by some amount, also specified by the threshold [5]. Choice of an effective threshold is the primary goal of wavelet shrinkage algorithms [11]. For this paper, thresholds proposed in [6] and [9] are used.

The version of *BayesShrink* described in [6], proposed a scale dependent threshold that balances the requirements of smoothing and edge preservation:

$$T(s) = \begin{cases} \sqrt{\frac{\log N_s}{2s}} \frac{\sigma_l^2}{\sigma_x} & \sigma_l^2 < W_s^2 \\ \max_u |W(u,s)|, & otherwise \end{cases} \quad (3)$$

$$\sigma_x = \sqrt{\max(W_s^2 - \sigma_l^2, 0)} \quad (4)$$

The threshold described by Equations (3) and (4) encodes the fact described in [6] that a near optimal threshold in the case of Gaussian noise, consists of a ratio of the noise variance to the signal variance. As described in [6] the additional, scale dependent factor multiplying the ratio in Equation (3), was added to overcome the fact that the ratio alone is less effective at reducing noise artifacts around edges than it is at reducing noise artifacts for large constant regions of the image.

To enhance the ability of the restoration algorithm to preserve textural properties, a multifractal threshold algorithm proposed in [9] is used. The basis for that algorithm is to utilize the scaling behavior of multifractal patterns that effectively model textural qualities. To accomplish this the authors show how to extend information obtained at coarse scales to estimate values at finer scales. Standard restoration methods, including wavelet shrinkage algorithms, are generally not effective at estimating values at fine scales, where the data is assumed to be dominated by noise.

As has been well documented in the literature, a useful measure of fractal or multifractal behavior is the Holder exponent [13] [14] [8], $\alpha$, a measure related to the local fractal dimension and which characterizes the local regularity of a function. To formally define the Holder exponent, first consider the definition of Holder continuity for a function $f$ from $R^N$ to $R$ at a point $x_0$ in the domain of $f$. For nonnegative real constants $C$ and $h$, $f$ is Holder continuous at $x_0$ if:

$$|f(x_0) - f(y)| \leq C\|x_0 - y\|^h \tag{5}$$

for all $y$ in the domain of $f$. Then the Holder exponent $\alpha(x_0)$ at $x_0$ is the supremum over all values $h$ for which Equation (5) holds. $\alpha(x_0)$ can be thought of as the exponent of the highest order polynomial that can be used to approximate $f(x)$ at $x_0$. The more irregular $f(x)$ is at $x_0$, the smaller is $\alpha(x_0)$.

In the case of data, such as images, estimation of the Holder exponent allows one to extrapolate to fine, noise-dominated scales, provided one assumes that the function underlying the generated data exhibits scaling behavior that extends from coarser scales to those finer scales.

A number of methods such as box counting [15] and large deviation methods [16] have been proposed for estimating the Holder exponent from data, usually as part of a larger multifractal formalism. Meyer [17] and Jaffard [13] realized that under mild regularity conditions an estimate of the local Holder exponent could be computed from wavelet coefficients as:

$$\alpha_n = \lim_{2^{-s} \to 0} \frac{-\log_2 |W(u,s)|}{s} \tag{6}$$

where we have written $\alpha_n = \alpha(u, s)$ for ease of notation below, noting that representation of both scale and position and coordinates, via a single index should not lead to any confusion given context. Using this realization, finer scale values of $\alpha_n$ can be estimated by extending a linear fit of $\log_2 |W(u,s)|$ vs. $s$ to small $s$. More sophisticated methods than simple linear regression for estimating $\alpha_n$, such as modulus maximum methods [18] and wavelet leader methods [19] that use more specific local information, have been developed. However, for the purposes of generating prior parameter functions, the straightforward linear regression method was found to be effective. Note also that for these linear fits an intercept, labeled $k_n$, is generated and will be used for determining the fine scale multifractal estimates.

For the method proposed in [9] a threshold, similar to that for *BayesShrink* is proposed. That is, the threshold, or critical scale as it is referred to in [9], is effectively determined by the scale at which the noise amplitude becomes comparable to the signal amplitude. Since the primary concern is detecting multifractal structure at fine scales, only the values to be chosen below this critical scale are specified. So, for scales below the critical scale, and characterized in terms of the proposed use as a prior parameter function for the mean:

$$W_p(u,s) = \min\left(|W_l(u,s)|, 2^{k_n - j\left(\alpha_n + \frac{1}{2}\right)}\right) \text{sgn}(W_l(u,s)) \tag{7}$$

Equation (7) suggests that at fine scales, smaller values, inferred by using estimated Holder exponents from larger scales, should be retained in favor of larger values that, on average, would indicate only the presence of noise.

There are a number of ways that the *BayesShrink* and multifractal estimation methods described above could be incorporated into a Bayesian restoration algorithm in the wavelet domain. But as a first step a direct incorporation seemed most illuminating in terms of testing the efficacy of *WIMP* and understanding its behavior. To accomplish this, the prior mean and variance functions defined in Equations (8) and (9) below and illustrated in Figure (1) are applied at scales above the critical scale $s_c$:

$$W_p(u,s) = \begin{cases} T(s)\Lambda^2 & if\ W_l(u,s) < -T(s) \\ 0 & if\ -T(s) \le W_l(u,s) < T(s) \\ -T(s)\Lambda^2 & if\ W_l(u,s) \ge T(s) \end{cases} \tag{8}$$

$$\sigma_p^2(u,s) = \begin{cases} \Lambda^2 \sigma_l^2 & if\ W_l(u,s) < -T(s) \\ \varepsilon & if\ -T(s) \le W_l(u,s) < T(s) \\ \Lambda^2 \sigma_l^2 & if\ W_l(u,s) \ge T(s) \end{cases} \tag{9}$$

and the prior mean and variance functions defined in Equations (10) and (11) below and illustrated in Figure (2) are applied on scales finer than the critical scale $s_c$:

$$W_p(u,s) = \begin{cases} -2^{k_n - j(\alpha_n + 1/2)} & if\ W_l(u,s) < -2^{k_n - j(\alpha_n + 1/2)} \\ 0 & if\ -2^{k_n - j(\alpha_n + 1/2)} \le W_l(u,s) < 2^{k_n - j(\alpha_n + 1/2)} \\ 2^{k_n - j(\alpha_n + 1/2)} & if\ W_l(u,s) \ge 2^{k_n - j(\alpha_n + 1/2)} \end{cases} \tag{10}$$

$$\sigma_p^2(u,s) = \begin{cases} \varepsilon & if\ W_l(u,s) < -2^{k_n - j(\alpha_n + 1/2)} \\ \Lambda^2 \sigma_l^2 & if\ -2^{k_n - j(\alpha_n + 1/2)} \le W_l(u,s) < 2^{k_n - j(\alpha_n + 1/2)} \\ \varepsilon & if\ W_l(u,s) \ge 2^{k_n - j(\alpha_n + 1/2)} \end{cases} \tag{11}$$

The parameters $\varepsilon$ and $\Lambda^2$ are introduced to allow the prior mean and variance parameter functions to accurately mimic *BayesShrink* and the multifractal estimates. The particular values of $\varepsilon$ and $\Lambda^2$ are only important in their relationship to $\sigma_l^2$ in terms of choices that maintain the conditions: $\varepsilon \ll \sigma_l^2$ and $\Lambda^2 \gg \sigma_l^2$. One can check that these conditions accurately mimic *BayesShrink* and the small scale multifractal estimates by plugging the values specified for the parameter functions in Equations (8), (9), (10), and (11) into Equation (1) for the MAP estimate of the wavelet coefficient at position $u$ and scale $s$.

The only unspecified parameter that requires much discussion is the critical scale, $s_c$. The basic intuition, based on the arguments in [9], that the critical scale $s_c$. occurs for $\frac{\sigma_l^2}{W_s^2} 1$, implies that this should provide the test for $W_s^2$ that determines the critical $s_c$. But it could be that on small scales there is a single large coefficient that skews the estimate of $W_s^2$ for that scale. Another issue is that, from a methodological point of view, it's necessary that there are enough scales coarser than $s_c$. for proper estimation of the Holder exponents at scales less than or equal to $s_c$.. If that isn't the case, $s_c$. determined by the above criteria might turn out to be too coarse a scale to serve as the critical scale.

Given these caveats and other concerns raised in [9], those authors argue that the method they propose is most suitable for highly-sampled, one-dimensional signals. While in principle the methods should be effective for image restoration applications as well, the authors argue in [9] that the resolution in typical image processing applications is too low for the method to be effective. But as will be seen in the examples below, even at moderate resolution, the method is effective for use with *WIMP*, i.e. when used to generate prior parameter means and variances. But given the modest resolutions typically dealt with in image processing applications, and therefore used for the examples in this paper, the algorithm was most effective when using the finest scale as the critical scale $s_c$. That is, for the examples provided here, and more for practical than theoretical reasons, only the finest scale wavelet coefficients were estimated using the multifractal estimation method (via Holder exponents).

To summarize the steps of the *WIMP* restoration algorithm:

1. Determine parameters $\varepsilon$ and $\Lambda^2$, and the critical scale $s_c$ (based on image resolution; as described above for the examples in this paper the critical scale is the finest wavelet scale).

2. Choose a wavelet basis and transform the image to wavelet space.

3. Estimate the Holder exponents for all indices in wavelet space $s, u$ for which $s$ is at or finer than the critical scale $s_c$ – i.e. based on linear regression of $log_2|W(u,s)|$ on $s$.

4. Use the conjugate Gaussian formula, Equation (1), and the prior parameter functions specified in Equations (8) and (9) (effectively implementing *BayesShrink* in the current version of algorithm) to obtain the MAP estimate coefficients at scales coarser than the critical scale $s_c$.

5. Use the conjugate Gaussian formula, Equation (1), and the prior parameter functions specified in Equations (10) and (11), and the Holder exponents (and intercepts) obtained in step 3 (effectively implementing the multifractal estimation method) to obtain the MAP estimate coefficients for scales at or finer than the critical scale $s_c$.

6. Perform the inverse wavelet transform on the MAP estimated coefficients to obtain the restored image.

As mentioned above, choice of a particular wavelet basis, most suitable for a given application is often an issue. For the examples studied in this paper there was some sensitivity to the choice of wavelet basis, but a more important concern was that the order of the wavelet be [20] small enough to provide enough coarse-grained scales above the critical scale for reasonably robust estimation of the fine-grained Holder exponents. In the examples presented in the paper second-order Daubechies wavelets, i.e. with 2 vanishing moments, were used.

Implementation of *WIMP* was performed in Python using the PyWavelets [21], SciPy [22], Numpy [23], Matplotlib [24], and sckit-image [25] libraries and the code is available on GitHub [20].

Note in passing that use of the likelihood mean and variance in the definitions of the prior parameter functions, so as to emulate *BayesShrink* and multifractal estimation, obviously violates the letter of Bayesian analysis. It is nonetheless consistent with a number of empirical Bayes methods, including [6] which have provided various justifications for generalization of Bayesian analysis. In this respect it should be noted that though the likelihood values are used to determine the values of the prior parameter functions, the form of the thresholds that determine how that is accomplished are predetermined. Note further that, under the, perhaps questionable, assumption of the independence of modes, using wavelet coefficients to estimate wavelet coefficients at other scales is equivalent to assuming some degree of regularity or smoothness in the image.

## Examples

As a first example a simulated image that was expected to highlight the strengths and weaknesses of the proposed restoration method as against straight wavelet shrinkage is presented. Since *BayesShrink* and *VishuShrink* [5], which uses a universal threshold, are considered state of the art in terms of mainstream shrinkage restoration methods, they are compared to *WIMP*. As *BayesShrink* is the basis for the coarse-grained analysis for *WIMP*, this comparison is also useful for seeing what, if anything, the proposed modification adds to direct application of *BayesShrink*.

To generate the simulated image, a long-standing model of multifractal, fractional Brownian motion, was used [14] [26] [27]. A number of authors realized that fractional Brownian motion provided a useful model of the correlated behavior observed in time series of values, such as velocities, associated with turbulent fluids. This eventually led to the understanding that associated multifractal parameters, such as the Holder exponent, were useful for studying the correlated, scaling behavior of such systems. Another measure, the Hurst exponent, related to the Holder exponent, but characterizing global regularity rather than local regularity, is the parameter usually used to classify fractional Brownian motion.

The simulated image presented here was generated with a background consisting of a 2D fractional Brownian motion pattern, with Hurst exponent = 0.33, using the package [28], and based on the algorithms outlined in [26] [27]. Superimposed on this background are a set of semi-transparent to opaque figures with a variety of boundary structures (sharp to diffuse) used to probe the strengths and weaknesses of the different restoration methods.

Figure (3) shows the original, simulated image, the image with added noise, and restoration using *WIMP*, *BayesShrink*, and *VishuShrink*.

To compare the methods on a more standard image, often used for comparison of methods, figure (4), shows the cameraman image, the image with added noise, and restoration using using *WIMP*, *BayesShrink*, and *VishuShrink*.

# Discussion

The goal of the project described in this paper was to extend the previous work for Bayesian image processing in Fourier space [4] to more general representation of images in orthogonal representations. In particular a specific, initial, goal was to extend the restoration method to the wavelet domain to take advantage of the fact that wavelets provide the ability to focus on restoring more local properties of images than do Fourier modes. In addition, it was desired to incorporate insights that have been developed for understanding image properties in wavelet bases, specifically those associated with wavelet shrinkage algorithms and wavelet based multifractal analysis.

*WIMP*, developed to achieve the above aims, performed well with respect to the first example above. It was able to substantially restore regions of the 2D fractional Brownian motion generated background where *SureShrink* and *BayesShrink* were not. Not surprisingly *SureShrink* and *BayesShrink* performed slightly better at restoring regions around edges of certain objects in the image. This was to be expected because for those algorithms noise at finer scales is virtually ignored. The cost of focusing on the finer scales to enhance multifractal structure is that effects like noise induced fluctuations around the edges are likely to also be enhanced.

The cameraman image used for the second example which is commonly used to compare restoration algorithms where there is interest in the reconstruction of both edges and smooth regions in images, should play to the strengths of the shrinkage algorithms over *WIMP*. That was true for the edges, likely for the reasons stated above for the first example. But perhaps surprisingly, *WIMP* did much better at restoring the large smooth areas of the image, reducing the noise more effectively than either of the shrinkage algorithms. This points out an interesting property of the Holder exponents: They are a general measure of local regularity, though they are typically utilized as a parameter to characterize irregularity (smaller Holder exponent values) such as is the case for multifractal sets. But as this example illustrates, Holder exponent values are also able to effectively characterize regularity and consequently allow for the extrapolation of smooth regions typically dominated by noise.

Given the results of these examples it appears that the strength of *WIMP*, over other algorithms, is clearly in the area of restoration of textural properties in images. But the examples show, that incorporation of *BayesShrink* for larger scales, means that not much is sacrificed in terms of edge resolution for *WIMP*, and preservation of smoothness in constant regions is actually enhanced.

*WIMP* would be indicated for use in projects such as medical imaging studies for which the global structures are relatively stable, particularly for co-registered images, and for which the goal is to distinguish subtle tissue properties via textural characteristics.

## Conclusions

A Bayesian image restoration framework that utilized insights from wavelet shrinkage and multifractal estimation methods (*WIMP*) was presented. It was shown to be highly effective at characterizing image texture relative to the commonly used wavelet shrinkage methods such as *BayesShrink* and *VishuShrink*. Though effective at resolving image texture the method was somewhat limited relative to the shrinkage methods with respect to clear identification of edges in the image. While the effective use of *BayesShrink* at coarse scales helped to mitigate this effect, future extensions of the method should investigate ways to incorporate more effective edge detection, though some trade off in terms of retaining effectiveness at texture recovery might be inevitable. But for certain imaging studies, e.g. those aimed at detecting subtle texture differences in medical images, the method, even as currently implemented, appears to offer clear advantages.

Given the effectiveness of *WIMP*, a natural question might be why not just use the multiple threshold method combining *BayesShrink* and the multifractal estimates without using the Bayesian restoration framework. The answer is that the Bayesian restoration framework allows for a number of easily implemented extensions of *WIMP*, whereas ways to extend the direct threshold methods aren't as obvious. For instance, implementing empirical methods for determination of the prior parameter functions, as was described above, is conceptually straightforward.

Empirically based extensions of *WIMP* would be straightforward to implement. For certain classes of images, empirical determination of proper prior parameter functions in the wavelet domain could be highly effective due to the increased ability to characterize local features. An example of this situation would be imaging of similar structures using fixed imaging protocols such as co-registered anatomical images based on fixed MRI protocols. In that case an empirically derived prior could be based on a database of such images by estimating the distribution at each $(s, u)$ pair.

Other obvious extensions would be to try to generate more robust Holder exponent estimates by utilizing modulus maximum [18] and wavelet leader [19] methods, as well as to do a more detailed exploration of the effect of using different wavelet bases, or a generalization to wavelet packets (though here the independence assumption might be problematic). Other investigators have argued that the effectiveness of the analysis of particular image classes can depend sensitively on the type of wavelet used. But a limitation to these extensions is that in most cases improvements due to the choice of wavelet class, and robustness of Holder estimates depend on having large data sets, e.g. high-resolution images. Given the moderate resolution typical of image data in many situations, however, the simpler methods, and low order wavelets, employed for *WIMP* may turn out to provide the most effective and efficient approach.

Use of the Gaussian conjugate prior in the wavelet basis was convenient in the case of assuming Gaussian noise, and in terms of leading to the simple closed form of the MAP estimate in equation (1). Generalization to other distributions is straightforward, as can occur in the case of different image acquisition methods with different noise characteristics. But cases for which one can find an appropriate conjugate prior, and a closed form for the MAP estimate are preferred in terms of computational efficacy. Regardless of whether this is possible, splitting the optimization problem over the posterior distribution, into a product of independent optimizations, means that method will be more efficient than optimization over the posterior in image space.

# Acknowledgements

Research reported in this manuscript was supported by the National Institute of Biomedical Imaging and Bioengineering of the National Institutes of Health under award number R01EB022055.

KY would like to thank Jeffrey Scargle and David Donoho for early mentoring in the way of wavelets [29], James Crutchfield for support and collaboration in an early exploration of multifractals [30], and Gregory Lee for helpful suggestions on use of the PyWavelets package [21].

# Figures

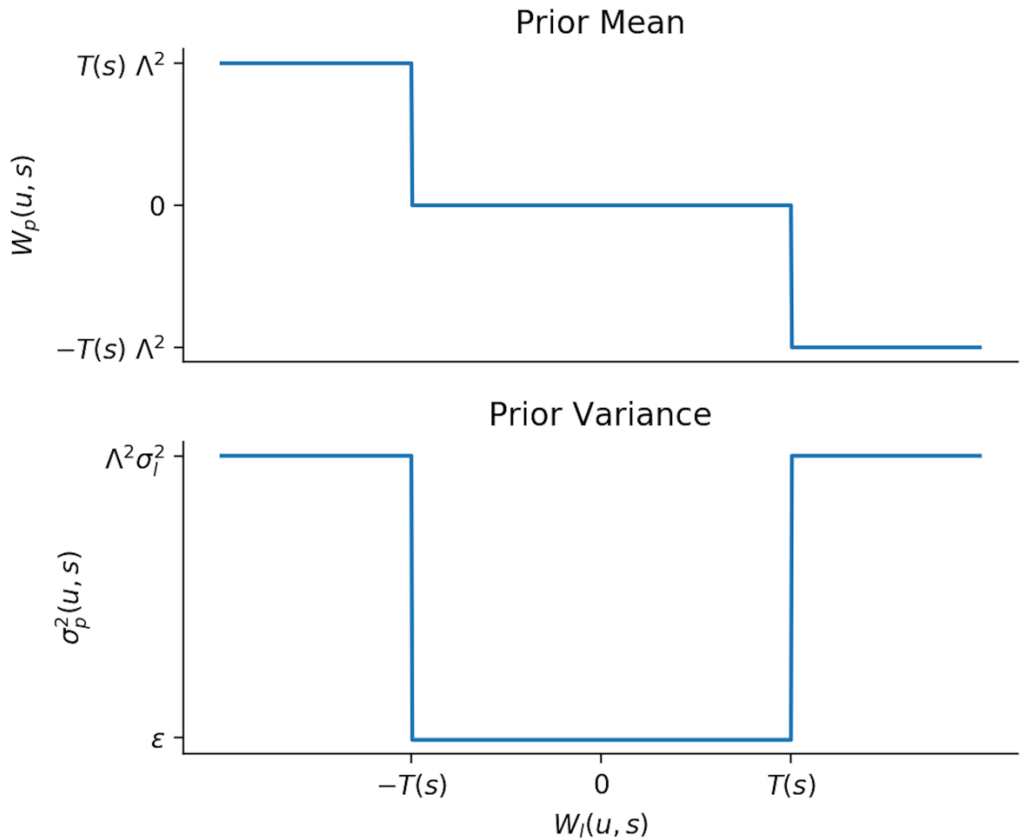

**Figure 1.**

Plot of the prior mean and prior variance parameter functions, $W_p(u, s)$ and $\sigma_p^2(u, s)$, to be applied at scales coarser than the critical scale. $W_p(u, s)$ and $\sigma_p^2(u, s)$ represent a direct implementation of *BayesShrink* in the Bayesian framework discussed in the text, where $T(s)$ is the scale dependent *BayesShrink* threshold and $\Lambda^2 \gg \sigma_l^2$ is a parameter that guarantees that the MAP estimate formula in Equation (1) effectively yields the standard *BayesShrink* coefficient estimates.

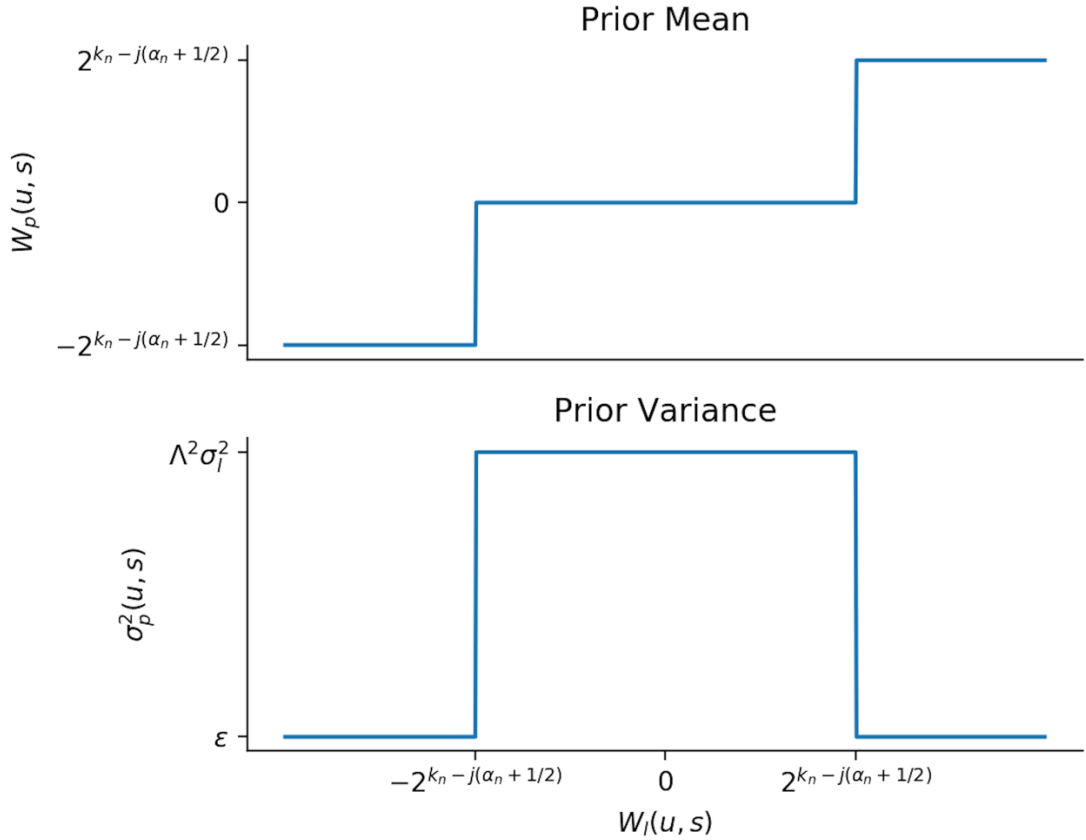

**Figure 2.**

Plot of the prior mean and prior variance parameter functions, $W_p(u,s)$ and $\sigma_p^2(u,s)$, to be applied at scales finer than the critical scale. $W_p(u,s)$ and $\sigma_p^2(u,s)$ represent a direct implementation of the Holder exponent based, multifractal estimate of coefficients described in [9], and in the Bayesian framework discussed in the text via the MAP estimate formula in equation (1). The scale and location dependent value $2^{k_n - j(\alpha_n + \frac{1}{2})}$, obtained from the linear regression estimates of $\alpha_n$ (the Holder exponent) and $k_n$ (the intercept for the regression) at location $u$ and scale $s$, serves as both a threshold when compared with $\sigma_l^2$ and the estimated coefficient value (or not), based on the result of the threshold comparison.

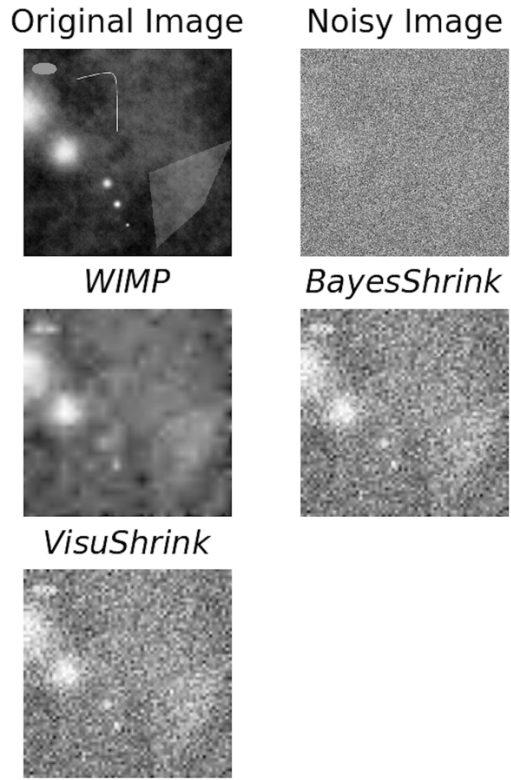

**Figure 3.**

Comparison of the *BayesShrink*, *VishuShrink*, and *WIMP* on the restoration of a simulated image with added noise. The simulated image consisted of a 2D fractional Brownian motion background with Hurst exponent 0.33, generated using the software package [28], to which a set of semi-transparent to opaque figures with a variety of boundary structures (sharp to diffuse) were superimposed.

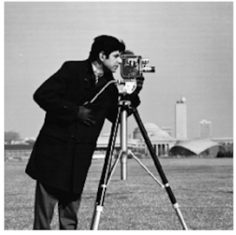
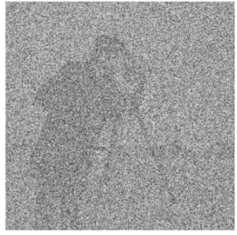
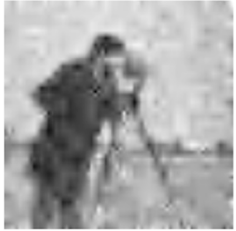
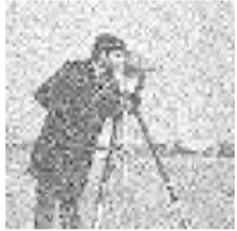
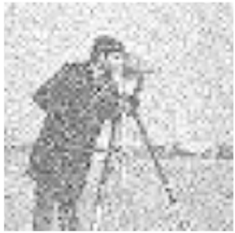

**Figure 4.**

Comparison of the *BayesShrink*, *VishuShrink*, and *WIMP* on the restoration of the standard cameraman image with added noise.